%% file: main.tex
\documentclass[conference]{IEEEtran}
\IEEEoverridecommandlockouts

\usepackage[mode=buildmissing]{standalone} 
\usepackage{amsmath}
\usepackage{bm}
\usepackage[nolist]{acronym}
\usepackage{amssymb}
\usepackage{comment}
\usepackage{graphicx}
\usepackage{color,colortbl}
\usepackage[dvipsnames]{xcolor}
\usepackage{cleveref}
\usepackage{verbatim}
\usepackage{enumerate}
\usepackage[shortlabels]{enumitem}
\usepackage[boxed,ruled]{algorithm2e}

\usepackage{cuted, nccmath}
\usepackage{dblfloatfix}
\usepackage{float} 
\usepackage{lipsum}
\usepackage{balance}
\usepackage{cite}
\usepackage{url}
\usepackage{tabularx}
\usepackage{boldline}
\usepackage{balance}    
\usepackage{mathtools} 
\usepackage{stmaryrd} 
\usepackage{graphicx}
\usepackage{pgfplots}
\usepackage{booktabs}
\usepackage[font={footnotesize}]{caption}
\usepackage{subfig} 
\usepackage{tikz}
\usetikzlibrary{arrows.meta}
\tikzset{every picture/.style={line width=0.6pt}}
\usetikzlibrary{calc,positioning}
\usetikzlibrary{arrows.meta,
                backgrounds,
                chains,
                fit,
                quotes}

\newcommand{\bbold}{\bm{b}}
\newcommand{\complexset}[2]{ \mathbb{C}^{#1 \times #2}  }
\newcommand{\Ngrid}{N_{\rm{grid}}}
\newcommand{\yy}{\bm{y}}
\newcommand{\yyhat}{\widehat{\yy}}
\newcommand{\AAb}{\boldsymbol{A}}
\newcommand{\hermit}{\mathrm{H}}
\newcommand{\trpose}{\mathrm{T}}

\newcommand{\norm}[1]{\left\lVert#1\right\rVert}
\newcommand{\atx}{\bm{a}_{\text{tx}}}
\newcommand{\arx}{\bm{a}_{\text{rx}}}
\newcommand{\conj}{ {\ast} }

\newcommand{\zz}{ \bm{z} }
\newcommand{\zr}{ \bm{z}_r }
\newcommand{\zc}{ {z}_c }

\newcommand{\m}{\bm{m}}
\newcommand{\llr}{\mathcal{L}}
\newcommand{\llrlog}{\mathcal{L}^{\rm{log}}}
\newcommand{\thetamax}{\theta_{\max}}
\newcommand{\thetamin}{\theta_{\min}}
\newcommand{\alphahat}{\widehat{\alpha}}
\newcommand{\thetahat}{\widehat{\theta}}

\newcommand{\pfa}{P_{\text{fa}}}
\newcommand{\SNR}{\text{SNR}}
\newcommand{\dB}{\text{dB}}

\newcommand{\hypz}{ \mathcal{H}_0 }
\newcommand{\hypone}{ \mathcal{H}_1 }
\newcommand{\hdet}{ \underset{\mathcal{H}_0}{\overset{\mathcal{H}_1}{\gtrless}} }
\newcommand{\etatilde}{\widetilde{\eta}}
\newcommand{\etabar}{\widebar{\eta}}

\newcommand{\etx}{ E_{\text{tx}} }

\newcommand{\degree}{^{\circ}}

\renewcommand{\Pr}{p}

\DeclarePairedDelimiter\abs{\lvert}{\rvert}%
\DeclarePairedDelimiter\absbigs{\big\lvert}{\big\rvert}%
\makeatletter
\newcommand*\rel@kern[1]{\kern#1\dimexpr\macc@kerna}
\newcommand*\widebar[1]{%
  \begingroup
  \def\mathaccent##1##2{%
    \rel@kern{0.8}%
    \overline{\rel@kern{-0.8}\macc@nucleus\rel@kern{0.2}}%
    \rel@kern{-0.2}%
  }%
  \macc@depth\@ne
  \let\math@bgroup\@empty \let\math@egroup\macc@set@skewchar
  \mathsurround\z@ \frozen@everymath{\mathgroup\macc@group\relax}%
  \macc@set@skewchar\relax
  \let\mathaccentV\macc@nested@a
  \macc@nested@a\relax111{#1}%
  \endgroup
}
\makeatother

\begin{acronym}[ACRONYM]
\acro{AD}[AD]{autonomous driving}
\acro{ADAS}{advanced driver-assistance system}
\acro{AE}{autoencoder}
\acro{AoA}{angle-of-arrival}
\acro{AoD}{angle-of-departure}
\acro{CRB}{Cram\'{e}r-Rao bound}
\acro{ML}{machine learning} 
\acro{IQ}{in-phase/quadrature} 
\acro{SNR}{signal-to-noise ratio}
\acro{JRC}{joint radar and communication}
\acro{NN}{neural network}
\acro{BCE}{binary cross-entropy}
\acro{NLL}{negative log-likelihood}
\acro{PDF}{probability density function}
\acro{PSD}{power spectral density}
\acro{CCE}{categorial cross-entropy}
\acro{MIMO}{multi-input multi-output}
\acro{MISO}{multi-input single-output}
\acro{SER}{symbol error rate}
\acro{RMSE}{root mean squared error}
\acro{MSE}{mean squared error}
\acro{ISAC}{integrated sensing and communication}
\acro{OFDM}{orthogonal frequency-division multiplexing}
\acro{6G}{6th generation wireless systems}
\acro{ReLU}{Rectified Linear Unit}
\acro{CSI}{channel state information}
\end{acronym}

\begin{document}
\bstctlcite{IEEEexample:BSTcontrol}

\title{End-to-End Learning for Integrated\\ Sensing and Communication}
\author{Jos\'{e} Miguel Mateos-Ramos\IEEEauthorrefmark{1}, Jinxiang Song\IEEEauthorrefmark{1}, Yibo Wu\IEEEauthorrefmark{1}\IEEEauthorrefmark{2}, Christian H\"{a}ger\IEEEauthorrefmark{1}, \\Musa Furkan Keskin\IEEEauthorrefmark{1}, Vijaya Yajnanarayana\IEEEauthorrefmark{3}, Henk Wymeersch\IEEEauthorrefmark{1}\\
\IEEEauthorrefmark{1}Department of Electrical Engineering, Chalmers University of Technology, Sweden \\
\IEEEauthorrefmark{2}Ericsson Research, Sweden, \IEEEauthorrefmark{3}Ericsson Research, India
}

\maketitle

\begin{abstract}
Integrated sensing and communication (ISAC) aims to unify radar and communication systems through a combination of joint hardware, joint waveforms, joint signal design, and joint signal processing. At high carrier frequencies, where ISAC is expected to play a major role, joint designs are challenging due to several hardware limitations. Model-based approaches, while powerful and flexible, are inherently limited by how well the models represent reality. Under model deficit, data-driven methods can provide robust ISAC performance. We present a novel approach for data-driven ISAC using an auto-encoder (AE) structure. The approach includes the proposal of the AE architecture, a novel ISAC loss function, and the training procedure. Numerical results demonstrate the power of the proposed AE, in particular under hardware impairments. 
\end{abstract}
\IEEEoverridecommandlockouts
\IEEEpeerreviewmaketitle

\section{Introduction} \label{sec:introduction}
Progressive generations of mobile communication systems have moved up in carrier frequency to unlock ever larger bandwidths, starting with 5G in the mmWave band and 6G envisioned to operate above 100 GHz \cite{tataria20216g,matthaiou2021road,saad2019vision}. 
The combination of large bandwidths and large arrays is reminiscent of high-resolution radar, available, e.g., to support \ac{AD} and \ac{ADAS} applications in moderns vehicles \cite{patole2017automotive}. This observation has led to the introduction of \ac{ISAC}, where the same spectrum is used for both radar-like sensing and high-rate communication \cite{chiriyath2017radar,tan2021integrated,liu2021integrated,rahman2020enabling,wymeersch2021integration}. 

According to \cite{liu2021integrated}, ISAC's history can be traced back in the radar community to the 1960s, an example of which is the missile range instrumentation radar \cite{ISAC_1963}.
In the communication community, ISAC has only recently found traction, after the introduction of \ac{OFDM} radar \cite{sturm2011waveform}. Unlike pulsed or continuous wave radars, OFDM radars are resilient to wireless channels due to the inherent frequency diversity which enhances the sensing performance \cite{sen-2010-adapt-ofdm}. ISAC systems can be developed in a number of ways, including (approximately) orthogonal  designs (in time \cite{han201224,kumari2021adaptive}, frequency \cite{aydogdu2019radchat}, or space \cite{liu2018mu,barneto2021beamformer}) and joint waveforms (referred to as unified designs in \cite[Table III]{liu2021integrated}). Joint waveforms are attractive from an efficiency point of view in monostatic\footnote{The ISAC literature has mainly focused on monostatic sensing, since for bistatic or multistatic sensing a pilot signal is transmitted. Hence, waveform design problems are different than in the monostatic case. } sensing, as the entire communication signal can be used for radar sensing and vice versa. 

The literature on joint waveforms for ISAC includes (i) communication waveforms used for sensing, e.g., \cite{sturm2011waveform,OFDM_Radar_Phd_2014}; (ii) sensing waveforms used for communication, e.g.,  \cite{hassanien2016signaling,scheiblhofer2015method}; and (iii) flexible designs that offer a trade-off between communication or sensing \cite{chiriyath2017radar,chen2021joint,liyanaarachchi2021joint,dokhanchi2021adaptive,johnston2021mimo,liu2018toward,liu2021cram,liu2021learning,zhang2018multibeam,OFDM_DFRC_TSP_2021}.
Existing approaches in the latter category differ in terms of the \emph{ISAC objective function} (e.g., radar and communication information rates \cite{chiriyath2017radar}, weighted radar peak-to-sidelobe level and communication  \ac{SNR} \cite{chen2021joint}, transmit power with interference constraints \cite{liyanaarachchi2021joint}, radar SNR under communication similarity constraint \cite{dokhanchi2021adaptive}, generalized radar metrics under communication error constraints \cite{johnston2021mimo}, communication interference subject to a communication similarity constraint \cite{liu2018toward}, radar \ac{CRB} under rate constraints \cite{liu2021cram}, communication rate under \ac{CRB} \cite{liu2021learning} and radar similarity \cite{OFDM_DFRC_TSP_2021} constraints) and the \emph{ISAC optimization variables} (e.g., power \cite{chiriyath2017radar,OFDM_DFRC_TSP_2021}, signal covariance \cite{chen2021joint}, beamformers \cite{liyanaarachchi2021joint,johnston2021mimo,liu2021cram,liu2021learning}, transmit sequences across antennas \cite{dokhanchi2021adaptive,liu2018toward}, weighted multibeams \cite{zhang2018multibeam}). 

Since the optimization problem in joint waveform design is often non-convex, approximate solution techniques are often applied, including those based on \ac{ML} \cite{liu2021learning}. Data-driven \ac{ML} methods are also useful under model deficits, e.g., to mitigate effects of array calibration errors, mutual coupling, power amplifier nonlinearity, quantization effects etc., which are expected to be prevalent in 6G \cite{wymeersch2021integration}. Hence, \ac{ML}-based designs 
are a promising alternative to conventional model-based approaches (see, e.g., \cite{simeone2018very,lang2020comprehensive}). In particular, end-to-end \acp{AE} \cite{o2017introduction}
are potentially well-suited for ISAC problems because they allow for the joint optimization of both the transmit waveforms as well as the communication and radar receivers. While \acp{AE} have been widely applied for communication \cite{aoudia2021waveform,cammerer2020trainable,zou2021channel,song2019learning} and radar \cite{jiang2019end,jiang2021joint,fuchs2020automotive} systems separately, AE-based designs have not been investigated in the ISAC literature.

  \begin{figure*}[h!]
        \centering
        \input{figures/jrc_bd}
        \caption{Block diagram of the \ac{ISAC} system model. The blocks highlighted in blue are implemented as trainable NNs as part of the proposed \ac{AE}. The radar receiver is assumed to be co-located with the transmitter, while the communication receiver is remote.}
        \label{fig:jrc_blocks} \vspace{-4mm}
    \end{figure*}
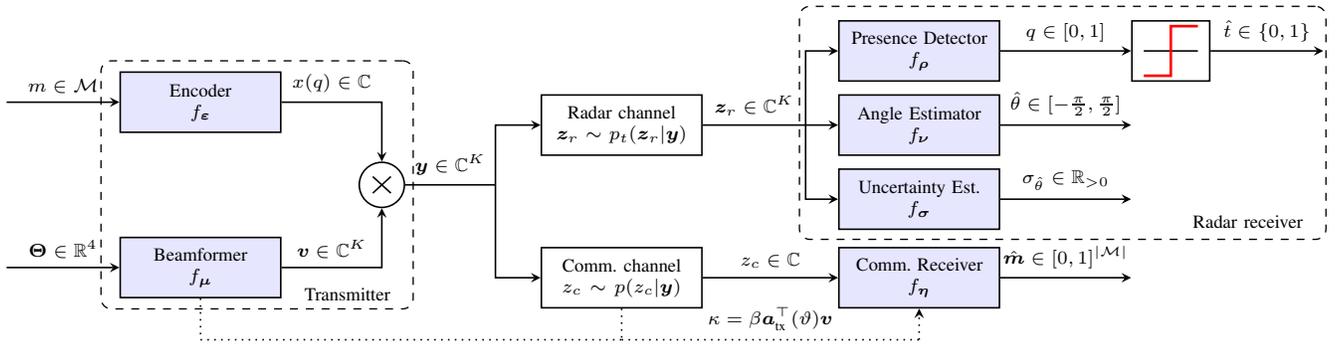
    
In this paper, we propose a novel \ac{AE} tailored to  \ac{ISAC}. We study a simplified single-target narrowband setting and generalize existing studies on end-to-end AE communication to the \ac{ISAC} setting. Our specific contributions are as follows: \emph{(i)} a novel \ac{AE} architecture to perform joint sensing and communication; \emph{(ii)} a novel loss function for radar sensing accounting for both target detection, target regression, and uncertainty quantification, which is subsequently combined with the standard communication \ac{CCE} loss; \emph{(iii)} a detailed performance comparison to the best known benchmarks, indicating similar performance; \emph{(iv)} a case study in the presence of hardware impairments, demonstrating the robustness of the proposed \ac{AE} over the model-based benchmarks.




\section{System Model}
    \label{sec:model}
    A block diagram of the considered system model is shown in Fig.~\ref{fig:jrc_blocks}.
    In the following, we first look at the radar and communication systems separately and then describe the model to perform the joint task of radar sensing and communication. 
    \subsection{Single-target MIMO Radar}
    \label{subsec:model:radar}
    We consider a \ac{MIMO} radar transceiver, which sends a complex signal $\bm{y} \in \mathbb{C}^K$ across $K$ antennas, subject to $ \mathbb{E}\{ \Vert \bm{y} \Vert^2 \} \le E_{\text{tx}}$. 
    %
    %
    %
    At the co-located radar receiver, a signal $\zr \sim p_t(\zr|\bm{y})$ across the $K$ receive antennas is observed, where $t\in \{0,1\}$ represents the absence or presence of a target, with $\Pr(t=1)=1/2$. In the absence of a target $\zr =\bm{n}$, while in the presence of a target
    \begin{align}
        \zr= 
        \alpha \bm{a}_{\text{rx}}(\theta)\bm{a}_{\text{tx}}^\top(\theta) \bm{y}+\bm{n}, 
        \label{eq:radar_scenario}
    \end{align}
    %
    where $\bm{n} \sim \mathcal{CN}(\bm{0}, N_0\bm{I}_K)$ with noise \acl{PSD} $N_0$, $[\bm{a}_{\text{tx}}(\theta)]_k=[\bm{a}_{\text{rx}}(\theta)]_k= \exp(-\jmath 2 \pi k d \sin (\theta ) / \lambda)$, with $d = \lambda / 2$. We further assume that $\alpha  \sim \mathcal{CN}(0,\sigma_r^2)$, following a  Swerling-1 model of the target, in which $\sigma_r^2$ captures the loss of power due to path loss and the target's radar cross section, and that the target (if present) is known to lie in a certain \ac{AoA} interval (equivalent to the \ac{AoD} interval)
    $\theta  \sim \mathcal{U}[\theta_{\min},\theta_{\max}]$, with $-\pi/2 \leq \thetamin \leq \thetamax \leq \pi/2$.

    
    
    The purpose of the radar receiver is to determine the probability $q\in [0,1]$ that  a target is present, and, if so, determine an estimate $\hat{\theta}$ of the \ac{AoA}  with an uncertainty estimate $\sigma_{\hat{\theta}}$. 
    
    
    \subsection{MISO Communication}
    \label{subsec:model:miso}
   The transmitter sends a message $m \in \mathcal{M}$, which should be mapped onto a constellation and precoded to achieve high \ac{SNR} at the receiver. We denote the transmitted signal across the $K$ antennas by $\bm{y}(m) = \bm{v} x(m)$ (again subject to $\mathbb{E}\{ \Vert \bm{y} \Vert^2 \} \le E_{\text{tx}}$), where 
    $\bm{v} \in \mathbb{C}^K$ is the transmit beamformer  and $x(m)\in \mathbb{C}$ is the mapping of the message in the \ac{IQ} plane. 
     We consider a remote receiver with one antenna. 
    %
    The observed signal is given by 
    \begin{align}
        \zc=\beta\bm{a}_{\text{tx}}^\top(\vartheta) \bm{y}(m)+{n},
        \label{eq:comm_scenario}
    \end{align}
    where the channel is modeled as Rayleigh, with $\beta  \sim \mathcal{CN}(0,\sigma_{c}^2)$, and the communication receiver is known to lie in a certain \ac{AoD} range $\vartheta \sim \mathcal{U}[\vartheta_{\min},\vartheta_{\max}]$, with $-\pi/2 \leq \vartheta_{\min} \leq \vartheta_{\max} \leq \pi/2$.
    
    The purpose of the communication receiver is to recover the transmitted message $m$. In order to focus on the core communication functionality, we assume that a pilot sequence has been sent prior to data transmission, so that the communication receiver has access to \ac{CSI} $\kappa=\beta\bm{a}_{\text{tx}}^\top(\vartheta) \bm{v}$ (see, e.g., \cite{qin2019deep} for ML-based \ac{CSI} estimation methods). 
    
    
    \subsection{Integrated Sensing and Communication}
    \label{subsec:model:isac}
    In the \ac{ISAC} setting, the goal of the transmitter is to design $\bm{y}(m)$ as well as the corresponding radar and communication receivers to jointly optimize communication and radar performance. The transmitter has knowledge of $\bm{\Theta} = [\theta_{\min}, \theta_{\max}, \vartheta_{\min}, \vartheta_{\max}]$, which accounts for the possible locations of the target and the communication receiver. Such a joint optimization must account for trade-offs between sensing and communication performance, as discussed in Section \ref{sec:introduction}. 
    
    Benchmark solutions for radar, communication, and \ac{ISAC} are deferred to Section \ref{subsec:results:bench}.
    

\section{ISAC End-to-end Learning}
\label{sec:method}
    
    To solve the \ac{ISAC} problem, we propose to use an end-to-end learning approach via a novel \ac{AE} architecture and associated loss functions, as described in the following. 

    \subsection{\ac{AE} Architecture}
    
    We implement each of the six highlighted blocks in Fig.~\ref{fig:jrc_blocks} as a feed-forward \ac{NN}. 
    In particular, we  express the encoder and beamformer as functions $f_{\bm{\varepsilon}}: \mathcal{M} \to \mathbb{C}$ and $f_{\bm{\mu}}: \mathbb{R}^4 \to \mathbb{C}^K$, respectively, where $\bm{\varepsilon}$ and $\bm{\mu}$ are the learnable parameters of each network. Similarly, the presence detector $f_{\bm{\rho}}: \mathbb{C}^K \to [0,1]$, angle estimator $f_{\bm{\nu}}: \mathbb{C}^K \to [-\pi/2, \pi/2]$, uncertainty estimator $f_{\bm{\sigma}}: \mathbb{C}^K \to \mathbb{R}_{>0}$, and communication receiver $f_{\bm{\eta}}: \mathbb{C} \to [0,1]^{|\mathcal{M}|}$ are a function of the learnable parameters $\bm{\rho}$, $\bm{\nu}$, $\bm{\sigma}$, and $\bm{\eta}$, respectively. The inputs and outputs to each \ac{NN} are shown in Fig.~\ref{fig:jrc_blocks}. The radar and communication channel blocks are both instantaneously differentiable, which means that they are differentiable under a realization of the random variables linked to them. This enables supervised end-to-end learning of all \acp{NN}, with training labels $[m,t,\theta]$.

    \subsection{Loss Functions}
    \subsubsection{Target Detection}
    \label{subsec:method:radar}
    The output from the detector is an estimate of the probability $q \in [0,1]$ that the target is present. During testing, a threshold can then be applied to $q$.  
    An appropriate metric for this type of estimation is the \ac{BCE} loss, defined as
    \begin{align}
        \mathcal{J}_{\text{TD}} (\bm{\varepsilon}, \bm{\mu}, \bm{\rho}) = -\mathbb{E}[t\log(q) + (1-t)\log(1-q)],
        \label{eq:method:radar_loss}
    \end{align}
    where the expectation is over the noise, the presence/absence of a target, 
    the radar channel gain, and the true target \ac{AoA}. 

    \subsubsection{Target Regression}
    \label{subsec:method:radarRegression}
    If a target is present, a regression loss can be used to assess how well the \ac{AE} determines the target's \ac{AoA}. Rather than simply using the \ac{MSE} $\mathbb{E}[|\hat{\theta}-\theta|^2]$, which only learns the target's \ac{AoA}, we propose to use the \ac{NLL} 
     \begin{align}
        \mathcal{J}_{\text{TR}} (\bm{\varepsilon}, \bm{\mu}, \bm{\rho},\bm{\sigma}) & = -\mathbb{E}[\log(p(\hat{\theta}|\theta)]\\
        & = \mathbb{E}\big[ \log(\sigma_{\hat{\theta}}) + \frac{1}{2\sigma_{\hat{\theta}}^2}\vert\theta - \hat{\theta}\vert^2 \big],
    \end{align}
    where we approximated the likelihood $p(\hat{\theta}|\theta)$ with a Gaussian density $\hat{\theta} \sim \mathcal{N}(\theta, \sigma_{\hat{\theta}}^2)$. Through this loss function, the receiver learns both the target's \ac{AoA} $\hat{\theta}$ and the  corresponding uncertainty $\sigma_{\hat{\theta}}$, which can be useful for subsequent processing.
    
    \subsubsection{Overall Radar Loss Function}
    \label{subsec:method:combined_radar_loss}
    Combining the detection and regression loss lead to a joint \ac{NLL} loss, proposed in \cite{pinto2021uncertainty}
    \begin{align}
        \mathcal{J}_{\text{NLL}} (\bm{\varepsilon}, \bm{\mu}, \bm{\rho}, \bm{\nu},\bm{\sigma}) &=\mathcal{J}_{\text{TD}} + \Pr(t=1)\mathcal{J}_{\text{TR}}. \label{eq:method:nll_loss}
    \end{align}

    \subsubsection{Communication Loss Function}
    \label{subsec:method:miso}
    We apply the widely used \ac{CCE} loss. Let $C = |\mathcal{M}|$, $\m^{\text{enc}} \in \{0,1\}^C$ be the one-hot encoding \cite{o2017introduction} of $m$ and $\hat{\m}\in [0,1]^C$ a $C$-dimensional probability vector. Then, the \ac{CCE} loss is 
    \begin{equation}
        \mathcal{J}_{\text{CE}} (\bm{\varepsilon}, \bm{\mu}, \bm{\eta}) = -\mathbb{E} \left[ \sum_{j=1}^C m^{\text{enc}}_j \log(\hat{m}_j) \right].
    \end{equation}
    
    \subsubsection{\ac{ISAC} loss}
    In order to combine the loss functions from the radar and communication transceivers, we consider a joint loss function as a linear combination of the individual losses
    \begin{equation}
        \mathcal{J}_{\text{ISAC}}(\bm{\varepsilon}, \bm{\mu}, \bm{\rho}, \bm{\sigma}, \bm{\nu}, \bm{\eta}) = \omega_r \mathcal{J}_{\text{NLL}} + (1-\omega_r) \mathcal{J}_{\text{CE}},
        \label{eq:method:jrc_loss}
    \end{equation}
    where $\omega_r \in [0,1]$ is a hyper-parameter to trade off radar performance for communication performance. 
    

\section{Results}
In this section, we describe the simulation parameters, the performance metrics, the benchmarks, and finally the simulation results with discussion. Cases without and with hardware impairments are considered. 

\subsection{Simulation Parameters and Metrics}
We set $|\mathcal{M}| = 4$, $K = 16$, and  $\mathbb{E}\{ \Vert \bm{y} \Vert^2 \} = 1$. The average \ac{SNR} in the communication is $\SNR_c = \sigma_c^2/N_0 = 20\,\dB$ (both for training and testing). The possible receiver locations lie in the range $(\vartheta_{\min}, \vartheta_{\max}) = (30\degree, 50\degree)$. The average \ac{SNR} in the radar model is $\SNR_r = \sigma_r^2 / N_0 = 0\,\dB$, and the target can be located in $(\thetamin, \thetamax) = (-20\degree, 20\degree)$.

To evaluate the communication performance, we use the \ac{SER} $\mathbb{E}[\Pr(\hat{m}\neq m)]$. To evaluate the radar performance, we use the detection probability $P_{\text{d}}=\Pr(\hat{t}=1| t=1)$, false alarm probability $P_{\text{fa}}=\Pr(\hat{t}=1| t=0)$, and \ac{RMSE}, $\sqrt{\smash[b]{\mathbb{E}[|\hat{\theta}-\theta|^2]}}$ (only when $\hat{t}=t = 1$, i.e., when a target is present and detected).

\subsection{Benchmarks}
\label{subsec:results:bench}
\subsubsection{Transmitter Benchmark}
As communication constellation, we use 4-QAM. For communication and radar beamforming vector, we use the approach from \cite{precoding_mmWave_JSTSP_2014,analogBeamformerDesign_TSP_2017}. In particular, given an certain angular range $[\thetamin, \thetamax]$ (i.e., either for communication or radar), 
let $\bbold \in \complexset{\Ngrid}{1}$ denote the desired beampattern at $\Ngrid$ angular grid locations $\{\theta_i\}_{i=1}^{\Ngrid}$, with
\begin{align}
    [\bbold]_i = 
    \begin{cases}
        \abs{\atx(\theta_i)}^2, &~~ \text{if} ~~\theta_i \in [\thetamin, \thetamax] \\
        0, &~~ \text{otherwise} 
    \end{cases} ~. \label{eq:beampatternDesign2}
\end{align}
 Let $\AAb = \left[ \atx(\theta_1) \, \ldots \, \atx(\theta_{\Ngrid}) \right] \in \complexset{K}{\Ngrid}$ the transmit steering matrix corresponding to those locations. 
Then, the beampattern synthesis problem can be formulated as 
$	\mathop{\mathrm{min}}\limits_{\yy} \norm{ \bbold - \AAb^\trpose \yy  }_{2}^2$, 
which has a simple closed-form least-squares (LS) solution $   \yy = (\AAb^\conj \AAb^\trpose)^{-1} \AAb^\conj \bbold$. 
After normalization, this provides us with 
a communication-optimal beam $\bm{y}_c$ and a radar-optimal beam $\bm{y}_r$. For the \ac{ISAC} scenario, we apply the approach from \cite{zhang2018multibeam}, 
and design the transmit \ac{ISAC} beam as
\begin{align}
    \bm{v}(\rho,\varphi) = \sqrt{E_{\text{tx}}}\frac{\sqrt{\rho} \bm{y}_r + \sqrt{1-\rho}e^{\jmath \varphi }\bm{y}_c}{\Vert\sqrt{\rho} \bm{y}_r + \sqrt{1-\rho}e^{\jmath \varphi }\bm{y}_c \Vert }. 
\end{align}
where $\rho \in [0,1]$ is a trade-off parameter and $\varphi \in [0,2 \pi)$ is a phase that can be used to provide coherency between multiple beams. Such a beam can then be optimized with respect to $\rho,\varphi$ in terms of different objectives \cite{kumari2021adaptive,zhang2018multibeam}. 
For our purpose, it is sufficient to sweep over $[\rho,\varphi]$ and for each value evaluate the \ac{SER}, \ac{RMSE}, detection and false alarm probabilities for the corresponding optimized communication and radar receiver benchmarks, detailed next.  

\subsubsection{Radar Detection Benchmark}
To derive a benchmark for radar detection, we resort to the maximum a-posteriori (MAP) ratio test (MAPRT) detector \cite{MAP_Detector_TSP_2021}, which generalizes the generalized likelihood ratio test (GLRT) detector \cite{glrt_2001} to the case with random parameters and thus can take into account the prior information on $\alpha$ and $\theta$. Details can be found in Appendix \ref{app:radarBenchmark}.

\subsubsection{Communication Receiver Benchmark}
We apply the maximum likelihood detector
\begin{align}
    \hat{m}(\zc) = \arg\min_{m \in \mathcal{M}}\norm{ \zc - \beta\bm{a}_{\text{tx}}^\top(\vartheta) \bm{v} x(m)} ^2,
\end{align}
which minimizes the \ac{SER}. 

\subsection{AE Training}

In terms of the \ac{NN} architectures, Table \ref{tab:results:nn} shows the size of the layers in each network, as well as the activation functions for the output layer. 
The activation function for the hidden layers is the \ac{ReLU} function. 
Complex-valued inputs are converted to real-valued by concatenating their real and imaginary parts. 
In the transmitter, after computing $\bm{y}$, we apply a normalization layer, which scales the transmitted signal to meet the power constraint, as proposed in \cite{o2017introduction}. To train the AE, we employed the widely used Adam optimizer \cite{kingma2014adam} with learning rate $0.01$ and mini-batch size $10000$. 
The data samples in each mini-batch are drawn independently from the corresponding distribution (source or channel). Thus, no data is reused between training and testing, preventing overfitting issues. We utilized a total of 20 million samples to train each \ac{NN}. 

\begin{table}[t]
\centering
\caption{Summary of the \ac{NN} architectures.}
\label{tab:results:nn}
\begin{tabular}{@{}cccc@{}}
\toprule
Network                & Input layer     & Hidden layers   & Output layer      \\ \midrule
Encoder $f_{\bm{\varepsilon}}$ & $|\mathcal{M}|$ & $(K, K, 2K)$   & 2 (linear) \\
Beamformer $f_{\bm{\mu}}$         & 4               & $(K, K, 2K)$   & $K$ (linear) \\
Presence det.~$f_{\bm{\rho}}$        & $2K$           & $(2K, 2K, K)$ & 1 (sigmoid)       \\
Angle est.~$f_{\bm{\nu}}$         & $2K$           & $(2K, 2K, K)$ & 1 (tanh)       \\
Uncertainty est.~$f_{\bm{\sigma}}$      & $2K$           & $(2K, 2K, K)$ & 1 (ReLU)          \\
Comm.~receiver $f_{\bm{\eta}}$        & $2$           & $(K, 2K, 2K)$ & $|\mathcal{M}|$ (softmax)  \\ \bottomrule
\end{tabular}
\end{table}

Given the losses in \eqref{eq:method:radar_loss}--\eqref{eq:method:jrc_loss}, we could train 
all six \acp{NN} from Table \ref{tab:results:nn} at the same time. However, we found that sequentially training the radar receiver \acp{NN} yielded better performance. We maintain the joint training structure of \eqref{eq:method:jrc_loss}, but with slight changes to $\mathcal{J}_{\text{NLL}}$. Namely, we first train $f_{\bm{\varepsilon}}, f_{\bm{\mu}}, f_{\bm{\eta}}, f_{\bm{\nu}}$ substituting $\mathcal{J}_{\text{NLL}}$ in \eqref{eq:method:jrc_loss} by a modified \ac{MSE} error, $\mathcal{J}_{\text{MSE}} = \Pr(t=1) \mathbb{E}[|\hat{\theta}-\theta|^2]$. 
Secondly, we freeze $\bm{\nu}$ and train $f_{\bm{\varepsilon}}, f_{\bm{\mu}}, f_{\bm{\eta}}, f_{\bm{\sigma}}$ using just the second term of \eqref{eq:method:nll_loss} in \eqref{eq:method:jrc_loss}. Finally, we freeze $\bm{\sigma}, \bm{\nu}$ and train $f_{\bm{\varepsilon}}, f_{\bm{\mu}}, f_{\bm{\eta}}, f_{\bm{\rho}}$ by substituting $\mathcal{J}_{\text{NLL}}$ with $\mathcal{J}_{\text{TD}}$.


\subsection{Simulation Results without Hardware Impairments}
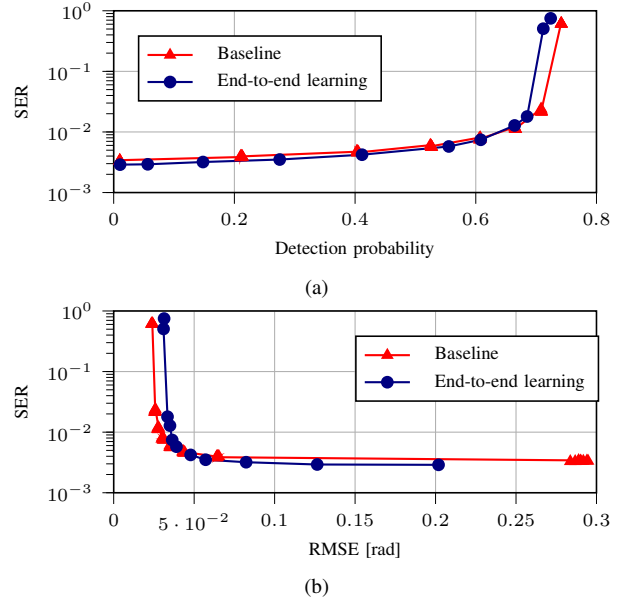
\begin{figure}[t]
    \centering
    \subfloat[]{\input{tikz/SER_vs_PD}}
    \vspace{-1mm}
    \subfloat[]{\input{tikz/SER_vs_RMSE}}
    \caption{Results (without hardware impairments) for a fixed empirical false alarm probability of $\pfa = 10^{-2}$, $\SNR_c = 20~\dB$, and $\SNR_r = 0~\dB$.}
    \label{fig:ser_vs_pd} \vspace{-4mm}
\end{figure}

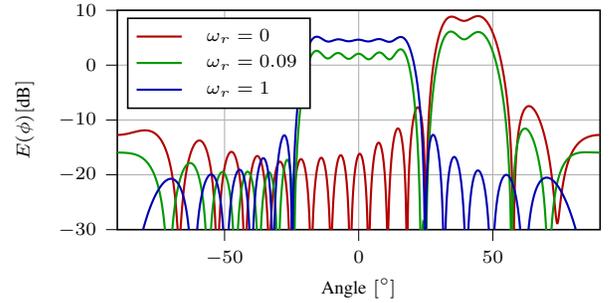
\begin{figure}[t]
    \centering
    \input{tikz/beams}
    \caption{Learned beampatterns (without hardware impairments) generated by the \ac{AE} for different values of the hyper-parameter $\omega_r$, where the communication receiver and the radar target reside, respectively, in the intervals $(30\degree, 50\degree)$ and $(-20\degree, 20\degree)$. The function $E(\phi) = \abs{\atx(\phi)^\top \yy}^2$ accounts for how much energy is transmitted in a certain direction.} \vspace{-4mm}
    \label{fig:beams}
\end{figure}

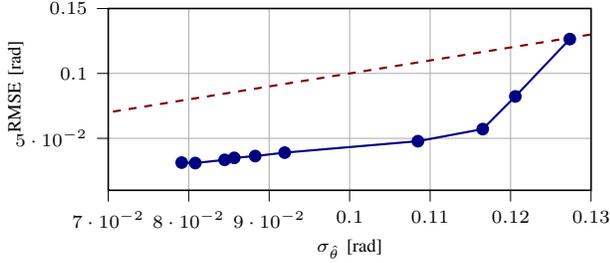
\begin{figure}[t]
    \centering
    \input{tikz/RMSE_vs_sigma}
    \caption{Results (without hardware impairments) of the \ac{RMSE} of the \ac{AoA} against the associated standard deviation $\sigma_{\hat{\theta}}$ for $\omega_r \in \{0.01, 0.014, 0.015, 0.03, 0.09, 0.15, 0.4, 0.6, 0.7, 1\}$. The dashed line shows $\ac{RMSE}=\sigma_{\hat{\theta}}$ as a reference. }
    \label{fig:rmse_vs_sigma}\vspace{-4mm}
\end{figure}

We show the \ac{ISAC} trade-off results in Fig.\,\ref{fig:ser_vs_pd} (a) (\ac{SER} vs.~detection probability) and Fig.\,\ref{fig:ser_vs_pd} (b) (\ac{SER} vs.~target \ac{RMSE}). 
In the test stage, we established a fixed false alarm probability of $\pfa = 10^{-2}$ and
computed the empirical value of $\pfa$ during testing to obtain these results.
Both figures indicate that the trade-off between 
radar and communication performance for the end-to-end learning approach based on different values of the hyper-parameter $\omega_r$ in \eqref{eq:method:jrc_loss} is close to the baseline.  This confirms that \ac{ML} approaches can perform as good as standard baselines for our particular scenario. The values of the hyper-parameters used in those simulations are $\omega_r \in \{0, 0.01, 0.014, 0.015, 0.03, 0.09, 0.15, 0.4, 0.6, 0.7, 1\}$.
We also observe a sharp degradation of communication performance when $\omega_r \to 1$, as the beamformer mainly illuminates the target and not the communication receiver, as seen in Fig.~\ref{fig:beams}. Conversely, when $\omega_r \to 0$, the beamformer illuminates the communication receiver, leading to severe radar performance degradation (i.e., low detection probability and high \ac{RMSE}). Nevertheless, there is a 'sweet spot' around $\omega_r \approx 0.09$, where both radar and communication achieve good performance, as the resulting beampattern points towards both angular sectors at the same time. 
Finally, in Fig.~\ref{fig:rmse_vs_sigma}, we assess the quality of the \ac{AoA} uncertainty estimate $\sigma_{\hat{\theta}}$. The \ac{RMSE} increases monotonically with $\sigma_{\hat{\theta}}$ as $\omega_r$ varies, though we slightly under-estimate the \ac{RMSE}. 



\subsection{Simulation Results under Hardware Impairments}
\begin{figure}[t]
    \centering
    \subfloat[]{\input{tikz/HI_SER_vs_PD}}
    
    \subfloat[]{\input{tikz/HI_SER_vs_RMSE}}
    \caption{Results (with hardware impairments) for a fixed empirical false alarm probability of $\pfa = 10^{-2}$, $\SNR_c = 20~\dB$, and $\SNR_r = 0~\dB$. }
    \label{fig:hi_ser_vs_pd} \vspace{-4mm}
\end{figure}
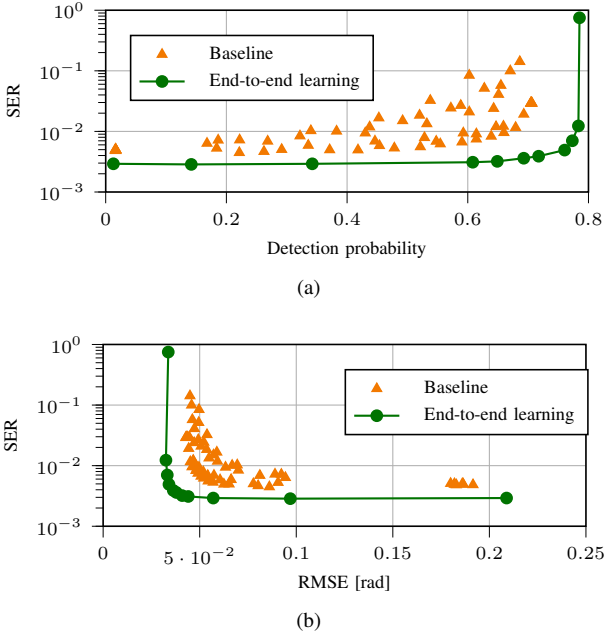

We now study the impact of a specific hardware impairment: the inter-element spacing, which up to now was assumed  to be  exactly $d = \lambda / 2$. Following \cite{yassine2020mpnet}, we apply a Gaussian perturbation, so that the distance between the $k$-th and $(k+1)$-th antenna elements is $d_k \sim_{\text{i.i.d.}} \mathcal{N}(\lambda/2, \sigma_{\lambda}^2)$. We set $\sigma_{\lambda} = \lambda/30$ and show the \ac{ISAC} trade-off results for a single realization of $d_k$ ($k=0,\ldots, K-2$) in Fig.~\ref{fig:hi_ser_vs_pd}. Note that the baseline assumes $d_k=\lambda/2$, $\forall k$. 
We observe that end-to-end learning can adapt to these hardware impairments,  whereas standard model-based approach without a perfect model incurs significant performance penalties (despite the very small deviations from the nominal model). In this case the hyper-parameter was selected to be $\omega_r \in \{0, 10^{-6}, 10^{-4}, 10^{-2}, 1.5\cdot 10^{-2}, 0.03, 0.05, 0.15, 0.4, 0.9, 1\}$.

\section{Conclusions}
In this work, we have proposed a novel end-to-end \ac{AE} approach for \ac{ISAC}, and we have compared the \ac{AE} performance with standard benchmarks for sensing and communications. Our results demonstrate that the trained AE performs close to the baseline. Moreover, we have shown the robustness of the proposed end-to-end learning approach to account for hardware impairments in the antenna array of the transmitter. 

Among possible future works, some natural extensions to this study include: (i) incorporate multiple targets to the sensing environment, (ii) use a \ac{MIMO} communication system, (iii) provide $\omega_r$ to the \ac{AE} input, (iv) learn across multiple angular ranges, and (v) make the channel more realistic towards 6G.

\section*{Acknowledgment}
{
{This work was supported, in part, by a grant from the Chalmers AI Research Center Consortium (CHAIR), by the European Commission through the H2020 project Hexa-X (Grant Agreement no. 101015956) and by the MSCA-IF grant
888913 (OTFS-RADCOM). The authors  gratefully acknowledge the feedback from Juliano Pinto and Lennart Svensson.}}

\appendices
\section{Radar Detection Benchmark} \label{app:radarBenchmark}
For the hypothesis testing problem where $\hypz$ and $\hypone$ denote the absence or presence of a target, the MAPRT corresponding to \eqref{eq:radar_scenario} can be written as \cite{MAP_Detector_TSP_2021}
\begin{align}\label{eq_maprt}
    \llr(\zr) = \frac{ \max_{\alpha, \theta, \yy} p(\alpha, \theta, \yy, \hypone \, \lvert \, \zr ) }{ p(\hypz \, \lvert \, \zr ) } \hdet \etatilde ~.
\end{align}
Notice that different from the Bayesian detector, we do not marginalize over $\alpha$ and $\theta$ in the MAPRT \cite{MAP_Detector_TSP_2021}. Applying the Bayes' theorem to \eqref{eq_maprt} yields
\begin{align}\label{eq_maprt2}
    \llr(\zr) &= \frac{ \max_{\alpha, \theta, \yy} p(  \zr  \, \lvert \, \alpha, \theta, \yy, \hypone) p(\alpha) p(\theta) p(\hypone) }{ p(\zr \, \lvert \,  \hypz ) p(\hypz) }  \hdet \etatilde ~.
\end{align}
Assuming $p(\hypz) = p(\hypone) = 1/2$ and taking the logarithm in \eqref{eq_maprt2}, we obtain 
\begin{align}\label{eq_maprt3}
    &\llrlog(\zr) = \frac{\norm{\zr}^2}{N_0} 
    \\ \nonumber
    &~~-   \min_{ \substack{\alpha, \theta \in [\thetamin, \thetamax] \\ \norm{\yy}^2 = \etx} } \left\{ \frac{\norm{ \zr - \alpha \, \bm{a}_{\text{rx}}(\theta)\bm{a}_{\text{tx}}^\top(\theta) \bm{y} }^2}{N_0} + \frac{\abs{\alpha}^2}{\sigma^2} \right\} \hdet \eta ~,
\end{align}
where $\llrlog(\zr) \triangleq \log \llr(\zr)$, $\eta \triangleq \log \etatilde + \log(\thetamax - \thetamin) +  \log(\pi \sigma^2) $, and the equality constraint on the transmit power is enforced to remove the ambiguity in estimating the channel gain $\alpha$. The optimal $\alpha$ in \eqref{eq_maprt3} can be computed for given $\theta$ and $\yy$ as
\begin{align}\label{eq_alphahat}
    \alphahat = \frac{ \yy^\hermit \atx^\conj(\theta) \arx^\hermit(\theta) \zr} { \norm{\bm{a}_{\text{rx}}(\theta)\bm{a}_{\text{tx}}^\top(\theta) \bm{y}}^2 + \frac{N_0}{\sigma^2} } = \frac{\yy^\hermit \atx^\conj(\theta) \arx^\hermit(\theta) \zr}{ K \abs{\atx^\top(\theta) \yy}^2 + \frac{N_0}{\sigma^2} } ~.
\end{align}
Plugging \eqref{eq_alphahat} back into \eqref{eq_maprt3} yields (after some algebraic manipulations)
\begin{align}\label{eq_maprt4}
    &\llrlog(\zr) =  \max_{ \substack{\theta \in [\thetamin, \thetamax] \\ \norm{\yy}^2 = \etx} } \frac{ \absbigs{ \atx^\top(\theta) \yy}^2 \absbigs{ \arx^\hermit(\theta) \zr}^2} { N_0 \left( K \absbigs{\atx^\top(\theta) \yy}^2 + \frac{N_0}{\sigma^2} \right)}    \hdet \eta ~.
\end{align}
From \eqref{eq_maprt4}, we can express the optimal $\yy$ as a function of $\theta$ as
\begin{align}\label{eq_yyhat}
    \yyhat = \sqrt{ \frac{\etx}{K} } \frac{ \atx^\conj(\theta) \arx^\hermit(\theta) \zr  }{ \absbigs{ \arx^\hermit(\theta) \zr} } ~.
\end{align}
Since $\absbigs{\atx^\top(\theta) \yyhat}^2 = \etx K$, inserting \eqref{eq_yyhat} into \eqref{eq_maprt4} yields the final detection test
\begin{align}\label{eq_maprt_etabar_no_y}
    \absbigs{ \arx^\hermit(\thetahat) \zz }^2   \hdet \etabar 
\end{align}
for some threshold $\etabar$ set to ensure a given false alarm probability, where $\thetahat \triangleq \arg \max_{\theta \in [\thetamin, \thetamax]} \absbigs{ \arx^\hermit(\theta) \zz}^2$.

\balance

\bibliographystyle{IEEEtran}
\bibliography{ref}

\end{document}

%% file: figures/jrc_bd.tex
\begin{tikzpicture}[font=\scriptsize, >=stealth,nd/.style={draw,circle,inner sep=0pt,minimum size=5pt}, blkCH/.style={draw,minimum height=0.8cm,text width=1.9cm, text centered},  blk/.style={draw,minimum height=0.8cm,text width=1.9cm, text centered}, rounded/.style={circle,draw,minimum size=0.6cm, text centered},  x=0.6cm,y=0.55cm]
\tikzset{mult_fig/.pic={
\draw (-1,-1)--(1,1);
\draw (-1,1)--(1,-1);
}}
\tikzset{threshold_fig/.pic={
\draw (-1,0)--(1,0);
\draw[red,line width=1.0 pt] (-1,-1)--(0,-1)--(0,1)--(1,1);
}}

\path		
	(-2,2)coordinate[](q){} 
	(-2,-2)coordinate[](theta){} 
	node(Encoder)[blk, fill=blue!10, right=2.5 of q]{Encoder \\ $f_{\bm{\varepsilon}}$}
	node(Beamformer)[blk, fill=blue!10, right=2.5 of theta]{Beamformer \\ $f_{\bm{\mu}}$}
	node(mult) [rounded, right=3.5 of $(Encoder)!0.5!(Beamformer)$]{}
	node[blk,rounded corners,fill opacity=0, dashed, line width=0.5, minimum width=4.15cm, minimum height=3.3cm](Tx)at (3.55,0.0){}
	node[](Tx) at(5.55,-2.65){Transmitter}  
	coordinate[right=2 of mult](tx_out){} 
	node(Channel)[blkCH, below right=1.5 and 1 of tx_out]{Comm. channel \\ $\zc \sim p(\zc|\bm{y})$}
	node(Receiver)[blk, fill=blue!10, right=3 of Channel]{Comm.~Receiver \\ $f_{\bm{\eta}}$}
	coordinate[right=2.9 of Receiver](q_hat){} 
	coordinate[below=1 of Beamformer](below_bf){} 
	coordinate[](below_ch) at (below_bf -| Channel){} 
	coordinate[](below_rec) at (below_bf -| Receiver){} 
	node(CH_radar)[blkCH, above right=0.7 and 1 of tx_out]{Radar channel \\ $\zr \sim p_t(\zr | \bm{y})$}
	coordinate[right=2.28 of CH_radar](channel_out){}  
	node(angle)[blk,fill=blue!10, ] at (channel_out -| Receiver){Angle Estimator \\  $f_{\bm{\nu}}$}	
	node(presence)[blk, fill=blue!10, above=0.3 of angle]{Presence Detector\\  $f_{\bm{\rho}}$}
	node(sigma)[blk, fill=blue!10, below=0.3 of angle]{Uncertainty Est.~\\ $f_{\bm{\sigma}}$}
	node(threshold)[blk,text width=0.8cm, right=2.9 of presence]{}
	coordinate[right=2.5 of threshold](m_hat){} 
	coordinate[right=2.9 of angle](theta_hat){} 
	coordinate[right=2.9 of sigma](sigma_hat){} 
	node[blk,rounded corners,fill opacity=0, dashed, line width=0.5, minimum width=7.0cm, minimum height=3.1cm](Tx)at (21.4,1.5){}
    node[](Tx) at(25.5,-0.9){Radar receiver}  
;
\draw[->] (q)--node[above]{$m \in \mathcal{M}$}(Encoder);
\draw[->] (theta)--node[above]{$\bm{\Theta} \in \mathbb{R}^4$}(Beamformer);
\draw[->] (Encoder)-|node[above,pos=0.25]{$x(q) \in \mathbb{C}$}(mult);
\draw[->] (Beamformer)-|node[above, pos=0.25]{$\bm{v} \in \mathbb{C}^K$}(mult);
\pic[scale=0.2](S) at (mult) {mult_fig};
\draw[-] (mult)--node[above]{$\bm{y} \in \mathbb{C}^K$}(tx_out);

\draw[->] (tx_out)|-(Channel);
\draw[->] (Channel)--node[above]{$\zc \in \mathbb{C}$}(Receiver);
\draw[->] (Receiver)--node[above]{$\hat{\m} \in [0,1]^{|\mathcal{M}|}$}(q_hat);

\draw[dotted] (Beamformer)--(below_bf);
\draw[dotted] (Channel)--(below_ch);
\draw[dotted] (below_bf)--(below_ch);
\draw[dotted] (below_ch)--node[above]{$\kappa = \beta \bm{a}_{\text{tx}}^\top(\vartheta)\bm{v}$}(below_rec);
\draw[dotted, ->] (below_rec)--(Receiver);

\draw[->] (tx_out)|-(CH_radar);
\draw[-] (CH_radar)--node[above]{$\zr \in \mathbb{C}^K$}(channel_out);
\draw[->] (channel_out)|-(presence);
\draw[->] (channel_out)|-(angle);
\draw[->] (channel_out)|-(sigma);
\draw[->] (presence)--node[above]{$q \in [0,1]$}(threshold);
\pic[scale=0.6](S) at (threshold) {threshold_fig};
\draw[->] (threshold)--node[above]{$\hat{t} \in \{0,1\}$}(m_hat);
\draw[->] (angle)--node[above]{$\hat{\theta} \in [-\frac{\pi}{2}, \frac{\pi}{2}]$}(theta_hat);
\draw[->] (sigma)--node[above]{$\sigma_{\hat{\theta}} \in \mathbb{R}_{>0}$}(sigma_hat);
\end{tikzpicture}

%% file: tikz/SER_vs_PD.tex
\begin{tikzpicture}[font=\scriptsize]
\begin{axis}[
width=8cm,
height=4.cm,
legend cell align={left},
legend style={
  fill opacity=1,
  draw opacity=1,
  text opacity=1,
  at={(0.05,0.5)},
  anchor=south west
},
log basis y={10},
tick align=outside,
tick pos=left,
x grid style={white!69.0196078431373!black},
xlabel={Detection probability},
xmajorgrids,
xmin=0, xmax=0.8,
xtick style={color=black},
y grid style={white!69.0196078431373!black},
ylabel={SER},
ymajorgrids,
ymin=1e-03, ymax=1,
ymode=log,
ytick style={color=black},
]
\addplot+[thick, color=red, mark=triangle*, mark options={scale=1}]
coordinates { 
(0.010145841466222235, 0.003328)
(0.009964897972238214, 0.003345)
(0.010131982135346778, 0.003359)
(0.009998959067324323, 0.003294)
(0.010191625777429141, 0.003303)
(0.01020438746700608, 0.003296)
(0.010333593172233216, 0.003357)
(0.010097213169165311, 0.003419)
(0.2135976940168545, 0.003862)
(0.2122777028040186, 0.00395)
(0.21272102612917804, 0.003954)
(0.21097288574176765, 0.004002)
(0.2102182363049667, 0.003775)
(0.21023968628706335, 0.00395)
(0.2115835982326716, 0.003865)
(0.212704527426506, 0.003953)
(0.4033767095366366, 0.00472)
(0.4030301878564442, 0.00483)
(0.4033341189912153, 0.004786)
(0.4053946545186328, 0.004687)
(0.4031497321928212, 0.004619)
(0.4037699483771388, 0.004515)
(0.40387572254913134, 0.00462)
(0.40472199301111467, 0.004582)
(0.5245671586295405, 0.006052)
(0.5254295656491637, 0.00603)
(0.5267152507643791, 0.005753)
(0.5264098244546972, 0.00559)
(0.5258728279304937, 0.005605)
(0.5263712278990562, 0.00569)
(0.5272446860860068, 0.005636)
(0.5254463011067428, 0.005786)
(0.6072506828006222, 0.008107)
(0.6057549966682229, 0.007806)
(0.6077617617617618, 0.007711)
(0.6072804633967365, 0.007643)
(0.6069223285648934, 0.007502)
(0.6061122751824657, 0.007396)
(0.6068953246851904, 0.007611)
(0.6068648768670972, 0.007694)
(0.6648033176019438, 0.012036)
(0.6642357958068478, 0.011842)
(0.6650073212291675, 0.011567)
(0.6647357841354589, 0.011227)
(0.6657230978071007, 0.011126)
(0.6663302636558597, 0.011094)
(0.6658357105873933, 0.011145)
(0.6657328964194373, 0.011356)
(0.7081928664835099, 0.024055)
(0.7083728524605909, 0.023306)
(0.7082178300992428, 0.022684)
(0.7092940180553026, 0.021759)
(0.7090095445093554, 0.021312)
(0.7080262931672151, 0.021522)
(0.7086911981724181, 0.021655)
(0.7085622216358417, 0.022229)
(0.7410996309224938, 0.611717)
(0.7416013589082316, 0.612141)
(0.7420619302300956, 0.611675)
(0.741598368848199, 0.611289)
(0.7415095132260949, 0.611692)
(0.742044922320653, 0.612637)
(0.7416053079537295, 0.612005)
(0.7412431003219184, 0.611877)
};\addlegendentry{Baseline};
\addplot +[thick, color=blue!50!black, solid, mark=*, mark options={scale=1,solid }] 
coordinates { 
(0.01057233446493549, 0.0028886666666666666)
(0.0564458578193948, 0.002925333333333333)
(0.14797558673614228, 0.003190333333333333)
(0.27508249883581437, 0.0035066666666666666)
(0.41149875902634697, 0.004212333333333333)
(0.5553303645301735, 0.005759666666666666)
(0.6082087011233491, 0.007424666666666666)
(0.6643656077033115, 0.012806999999999999)
(0.6850962922168957, 0.017977333333333335)
(0.7118785481553026, 0.5061080000000001)
(0.7240903501563727, 0.7497533333333334)
};
\addlegendentry{End-to-end learning}

\end{axis}

\end{tikzpicture}


%% file: tikz/SER_vs_RMSE.tex
\begin{tikzpicture}[font=\scriptsize]
\begin{axis}[
width=8cm,
height=4.cm,
legend cell align={left},
legend style={
  fill opacity=1,
  draw opacity=1,
  text opacity=1,
  at={(0.5,0.5)},
  anchor=south west
},
log basis y={10},
tick align=outside,
tick pos=left,
x grid style={white!69.0196078431373!black},
xlabel={RMSE [rad]},
xmajorgrids,
xmin=0, xmax=0.3,
xtick style={color=black},
y grid style={white!69.0196078431373!black},
ylabel={SER},
ymajorgrids,
ymin=1e-03, ymax=1,
ymode=log,
ytick style={color=black},
]
\addplot+[thick, color=red, mark=triangle*, mark options={scale=1}]
coordinates { 
(0.29181574291320883, 0.003328)
(0.29474467556928985, 0.003345)
(0.2940362499305172, 0.003359)
(0.283712442831598, 0.003294)
(0.29021399022965927, 0.003303)
(0.2867566320344358, 0.003296)
(0.29017088786926365, 0.003357)
(0.2888840628415276, 0.003419)
(0.06401867944472606, 0.003862)
(0.06461953176690834, 0.00395)
(0.06479680777250049, 0.003954)
(0.06534734854679802, 0.004002)
(0.06410240305441209, 0.003775)
(0.06464733620866156, 0.00395)
(0.06406270990153135, 0.003865)
(0.0639951919379509, 0.003953)
(0.0432184362725131, 0.00472)
(0.042932464109236124, 0.00483)
(0.04349680367840635, 0.004786)
(0.04297318778658529, 0.004687)
(0.04417369346286089, 0.004619)
(0.04338849955622828, 0.004515)
(0.04297977810217642, 0.00462)
(0.04379537904117178, 0.004582)
(0.03534477791883519, 0.006052)
(0.034901914219059416, 0.00603)
(0.03501236774687246, 0.005753)
(0.03547248416209334, 0.00559)
(0.03523520045525075, 0.005605)
(0.034755785889086135, 0.00569)
(0.035126214062967585, 0.005636)
(0.03499801081952197, 0.005786)
(0.03016653338292676, 0.008107)
(0.031189418337734958, 0.007806)
(0.029882219289764825, 0.007711)
(0.03070109624230676, 0.007643)
(0.03054000851679923, 0.007502)
(0.030885573803894267, 0.007396)
(0.030724731362933685, 0.007611)
(0.030962379978078162, 0.007694)
(0.02752762833131365, 0.012036)
(0.027964410568608752, 0.011842)
(0.027369013164309928, 0.011567)
(0.027289462031355113, 0.011227)
(0.027413490856472683, 0.011126)
(0.027803307084970873, 0.011094)
(0.027500258924565518, 0.011145)
(0.028042956957629407, 0.011356)
(0.025919400599424814, 0.024055)
(0.025566519287018363, 0.023306)
(0.025796829089586748, 0.022684)
(0.026181303827803194, 0.021759)
(0.025629584458631987, 0.021312)
(0.025623955579387733, 0.021522)
(0.025275419385958404, 0.021655)
(0.025832504456917076, 0.022229)
(0.02401513258330749, 0.611717)
(0.024257520207326626, 0.612141)
(0.02348900170680955, 0.611675)
(0.02376863470221471, 0.611289)
(0.023472832102933376, 0.611692)
(0.024030773171877934, 0.612637)
(0.023803955390976055, 0.612005)
(0.02409696087272915, 0.611877)
};\addlegendentry{Baseline};
\addplot +[thick, color=blue!50!black, solid, mark=*, mark options={scale=1,solid }] 
coordinates { 
(0.20185870841351028, 0.0028886666666666666)
(0.12645780451205887, 0.002925333333333333)
(0.08228758653704081, 0.003190333333333333)
(0.05714301291315937, 0.0035066666666666666)
(0.04781007563605368, 0.004212333333333333)
(0.03899031468788981, 0.005759666666666666)
(0.03636356031188276, 0.007424666666666666)
(0.03496881331503221, 0.012806999999999999)
(0.033443570197088525, 0.017977333333333335)
(0.030981801934043846, 0.5061080000000001)
(0.03133043406253949, 0.7497533333333334)
};
\addlegendentry{End-to-end learning}

\end{axis}

\end{tikzpicture}


%% file: tikz/beams.tex
\begin{tikzpicture}[font=\scriptsize]
\begin{axis}[
width=8cm,
height=4.5cm,
legend cell align={left},
legend style={
  fill opacity=1,
  draw opacity=1,
  text opacity=1,
  at={(0.02, 0.55)},
  anchor=south west
},
tick align=outside,
tick pos=left,
x grid style={white!69.0196078431373!black},
xlabel={Angle $[\degree]$},
xmajorgrids,
xmin=-90, xmax=90,
xtick style={color=black},
y grid style={white!69.0196078431373!black},
ylabel={$E(\phi) [\dB]$},
ymajorgrids,
ymin=-30, ymax=10,
ytick style={color=black},
]
\addplot+[thick, color=red!70!black, solid, mark=None, mark options={scale=1,solid }]table {tikz/beam0.txt}; \addlegendentry{$\omega_r = 0$}
\addplot+[thick, color=green!60!black, solid, mark=None, mark options={scale=1,solid }]table {tikz/beam_middle.txt}; \addlegendentry{$\omega_r = 0.09$}
\addplot+[thick, color=blue!70!black, solid, mark=None, mark options={scale=1,solid }]table {tikz/beam1.txt}; \addlegendentry{$\omega_r = 1$}


\end{axis}
\end{tikzpicture}


%% file: tikz/RMSE_vs_sigma.tex
\begin{tikzpicture}[font=\scriptsize]
\begin{axis}[
width=8cm,
height=4cm,
legend cell align={left},
legend style={
  fill opacity=1,
  draw opacity=1,
  text opacity=1,
  at={(0.02, 0.55)},
  anchor=south west
},
tick align=outside,
tick pos=left,
x grid style={white!69.0196078431373!black},
xlabel={$\sigma_{\hat{\theta}}$  [rad]},
xmajorgrids,
xmin=0.07, xmax=0.13,
xtick style={color=black},
y grid style={white!69.0196078431373!black},
ylabel={~~~~RMSE [rad]},
ymajorgrids,
ymin=0.01, ymax=0.15,
ytick style={color=black},
]
\addplot[thick, color=blue!50!black, solid, mark=*, mark options={scale=1,solid }]
coordinates {
(0.12736287157692014, 0.12645780451205887)
(0.12059077099204808, 0.08228758653704081)
(0.11653710410777479, 0.05714301291315937)
(0.10848729871798307, 0.04781007563605368)
(0.09192449584244937, 0.03899031468788981)
(0.08826441904391348, 0.03636356031188276)
(0.08565881077449769, 0.03496881331503221)
(0.08446511025762558, 0.033443570197088525)
(0.0808181343229115, 0.030981801934043846)
(0.07912737193526327, 0.03133043406253949)
}; 

\addplot[thick, color=red!50!black, dashed]
coordinates {
(0.2, 0.2)
(0.01, 0.01)
}; 



\end{axis}
\end{tikzpicture}


%% file: tikz/HI_SER_vs_PD.tex
\begin{tikzpicture}[font=\scriptsize]
\begin{axis}[
width=8cm,
height=4.cm,
legend cell align={left},
legend style={
  fill opacity=1,
  draw opacity=1,
  text opacity=1,
  at={(0.05,0.5)},
  anchor=south west
},
log basis y={10},
tick align=outside,
tick pos=left,
x grid style={white!69.0196078431373!black},
xlabel={Detection probability},
xmajorgrids,
xmin=0, xmax=0.8,
xtick style={color=black},
y grid style={white!69.0196078431373!black},
ylabel={SER},
ymajorgrids,
ymin=1e-03, ymax=1,
ymode=log,
ytick style={color=black},
]
\addplot+[only marks, mark=triangle*, mark options={scale=1, color=orange!95!black}]
coordinates { 
(0.0166950119050001, 0.00493)
(0.016829151350487456, 0.004962)
(0.01589493018956265, 0.00484)
(0.016624459764972798, 0.00502)
(0.016217706427150175, 0.005082)
(0.0164899835060131, 0.005012)
(0.016098761479821526, 0.00491)
(0.016966291236680763, 0.004818)
(0.2916112701473647, 0.005026)
(0.26226265506002194, 0.00466)
(0.22146212742138616, 0.004468)
(0.18387280136421497, 0.00528)
(0.16789864415738345, 0.006356)
(0.18638295826505716, 0.00718)
(0.22191772040322452, 0.00723)
(0.26839797396237586, 0.006894)
(0.45366672408336906, 0.005826)
(0.4180842219699755, 0.004946)
(0.37115092984276477, 0.00497)
(0.3357240860670984, 0.005884)
(0.32192546359910745, 0.008416)
(0.33998480425480865, 0.010268)
(0.38258957713425257, 0.010094)
(0.4301561183127985, 0.009412)
(0.5477393558592205, 0.006868)
(0.5214112437504745, 0.005572)
(0.4783376905795537, 0.005318)
(0.44617751583603443, 0.006922)
(0.4373761743465201, 0.01182)
(0.4526676902408598, 0.016688)
(0.49204243501888, 0.015014)
(0.5321777613866864, 0.013358)
(0.6142372962838325, 0.009088)
(0.5904016117233688, 0.00668)
(0.5544466862019823, 0.006222)
(0.5282647334891059, 0.007896)
(0.5201131124756171, 0.018464)
(0.5379122638488633, 0.032454)
(0.5719899792702156, 0.024302)
(0.6025132539561959, 0.020942)
(0.6591279283307516, 0.012168)
(0.638906227493158, 0.008318)
(0.6141777338513803, 0.00748)
(0.5926106972998313, 0.009414)
(0.588352748082153, 0.026804)
(0.6024640492809856, 0.084218)
(0.6274494911134741, 0.051464)
(0.6510941146382788, 0.040786)
(0.6928170064046556, 0.019212)
(0.6792717757247938, 0.011598)
(0.659837281070899, 0.00955)
(0.6468369566089671, 0.011964)
(0.6428574287784355, 0.023922)
(0.6548848816422512, 0.057438)
(0.6704002747296628, 0.099752)
(0.6862029692089436, 0.14191)
(0.7053890442638627, 0.029002)
(0.7063224990696911, 0.02931)
(0.7044315853502369, 0.029252)
(0.705917855087293, 0.029288)
(0.7042766151046406, 0.029114)
(0.7057285409959476, 0.029256)
(0.7052346996316634, 0.029614)
(0.7047644550312341, 0.029378)
};\addlegendentry{Baseline};
\addplot +[thick, color=green!45!black, solid, mark=*, mark options={scale=1,solid }] 
coordinates { 
(0.01275018286869401, 0.0029206666666666665)
(0.1418552125001115, 0.0028536666666666662)
(0.3423056453770505, 0.0029186666666666666)
(0.6082355744337673, 0.0031003333333333334)
(0.648931438592693, 0.003205333333333333)
(0.6927700384079131, 0.003609666666666667)
(0.7174211057642522, 0.0038956666666666666)
(0.7604757822227676, 0.004915666666666666)
(0.7732622287826832, 0.007012833333333333)
(0.7833057942485212, 0.012310866666666668)
(0.7851124460237446, 0.7502253333333333)

};
\addlegendentry{End-to-end learning}

\end{axis}

\end{tikzpicture}


%% file: tikz/HI_SER_vs_RMSE.tex
\begin{tikzpicture}[font=\scriptsize]
\begin{axis}[
width=8cm,
height=4.cm,
legend cell align={left},
legend style={
  fill opacity=1,
  draw opacity=1,
  text opacity=1,
  at={(0.5,0.5)},
  anchor=south west
},
log basis y={10},
tick align=outside,
tick pos=left,
x grid style={white!69.0196078431373!black},
xlabel={RMSE [rad]},
xmajorgrids,
xmin=0, xmax=0.25,
xtick style={color=black},
y grid style={white!69.0196078431373!black},
ylabel={SER},
ymajorgrids,
ymin=1e-03, ymax=1,
ymode=log,
ytick style={color=black},
]
\addplot+[only marks, mark=triangle*, mark options={scale=1, color=orange!95!black}]
coordinates { 
(0.1864758940174366, 0.00493)
(0.1863678700715465, 0.004962)
(0.19153340284754594, 0.00484)
(0.18599324831151667, 0.00502)
(0.17992481305617064, 0.005082)
(0.1819199067759009, 0.005012)
(0.18191377356948105, 0.00491)
(0.18288479671762378, 0.004818)
(0.07780582259306752, 0.005026)
(0.08026071677692338, 0.00466)
(0.08608187453490353, 0.004468)
(0.09060181861336573, 0.00528)
(0.09439012164342642, 0.006356)
(0.0922913166539941, 0.00718)
(0.08899422200313009, 0.00723)
(0.08114976049838203, 0.006894)
(0.06055101747726591, 0.005826)
(0.06252192236842656, 0.004946)
(0.06523484504714411, 0.00497)
(0.06627379054495781, 0.005884)
(0.07003061937386935, 0.008416)
(0.06945179942818017, 0.010268)
(0.06680969050924232, 0.010094)
(0.06337087438733995, 0.009412)
(0.0541680406924889, 0.006868)
(0.05435247111773104, 0.005572)
(0.05691881888013316, 0.005318)
(0.05738904828556919, 0.006922)
(0.0591986175277901, 0.01182)
(0.05887098927336546, 0.016688)
(0.05716174115258989, 0.015014)
(0.0548220964898725, 0.013358)
(0.04863070844957873, 0.009088)
(0.05067563913753694, 0.00668)
(0.05202678308824794, 0.006222)
(0.0518748775473393, 0.007896)
(0.053637475534657766, 0.018464)
(0.05377777136774198, 0.032454)
(0.052156210633406794, 0.024302)
(0.050714477048363515, 0.020942)
(0.046666768426687155, 0.012168)
(0.04801107645635914, 0.008318)
(0.0491226708105143, 0.00748)
(0.049588846813262054, 0.009414)
(0.04928036974312404, 0.026804)
(0.04964785832975372, 0.084218)
(0.049498696615112595, 0.051464)
(0.047231525611726645, 0.040786)
(0.04429608959487827, 0.019212)
(0.04515535730046, 0.011598)
(0.04601093200401677, 0.00955)
(0.04663522307867269, 0.011964)
(0.04712765921441298, 0.023922)
(0.046205703975854745, 0.057438)
(0.04578877983766149, 0.099752)
(0.044963767697792444, 0.14191)
(0.04280330572826886, 0.029002)
(0.04331679990320564, 0.02931)
(0.044633862943446075, 0.029252)
(0.043806461769122104, 0.029288)
(0.043670483981604315, 0.029114)
(0.044608780866227035, 0.029256)
(0.04298718499619507, 0.029614)
(0.04357793303164914, 0.029378)
};\addlegendentry{Baseline};
\addplot +[thick, color=green!45!black, solid, mark=*, mark options={scale=1,solid }] 
coordinates { 
(0.20890558142014765, 0.0029206666666666665)
(0.0969115133969952, 0.0028536666666666662)
(0.057020547181590164, 0.0029186666666666666)
(0.04407302003484073, 0.0031003333333333334)
(0.04092360416595118, 0.003205333333333333)
(0.037916297894691926, 0.003609666666666667)
(0.03633861070902976, 0.0038956666666666666)
(0.034085159276984286, 0.004915666666666666)
(0.033213247387028605, 0.007012833333333333)
(0.03251707412707041, 0.012310866666666668)
(0.033719531675926166, 0.7502253333333333)
};
\addlegendentry{End-to-end learning}

\end{axis}

\end{tikzpicture}
